\documentstyle[aps,epsf,multicol]{revtex}
\begin{document}
\title{Nanoscopic Tunneling Contacts on Mesoscopic Multiprobe Conductors}
\author{Thomas Gramespacher and Markus B\"uttiker}
\address{D\'epartement de physique th\'eorique, Universit\'e de Gen\`eve,
CH-1211 Gen\`eve 4, Switzerland}
\date{\today}
\maketitle
\begin{abstract}
We derive Bardeen-like expressions for the transmission probabilities
between two multi-probe mesoscopic conductors coupled by a weak tunneling
contact. We emphasize especially the dual role 
of a weak coupling contact as a current source and 
sink and analyze the magnetic field symmetry. 
In the limit of a point-like tunneling contact the transmission
probability becomes a product of local, partial density of states 
of the two mesoscopic conductors. We present expressions for 
the partial density of states in terms of functional derivatives 
of the scattering matrix with respect to the local potential and in terms 
of wave functions. We discuss voltage measurements and resistance measurements
in the transport state of conductors.
We illustrate the theory for the simple case of a scatterer 
in an otherwise perfect wire. In particular, we investigate the development
of the Hall-resistance as measured with weak coupling probes.
\end{abstract}
\begin{multicols}{2}
\narrowtext
\section{Introduction}
\begin{figure}
\narrowtext
\epsfysize3cm
\epsffile{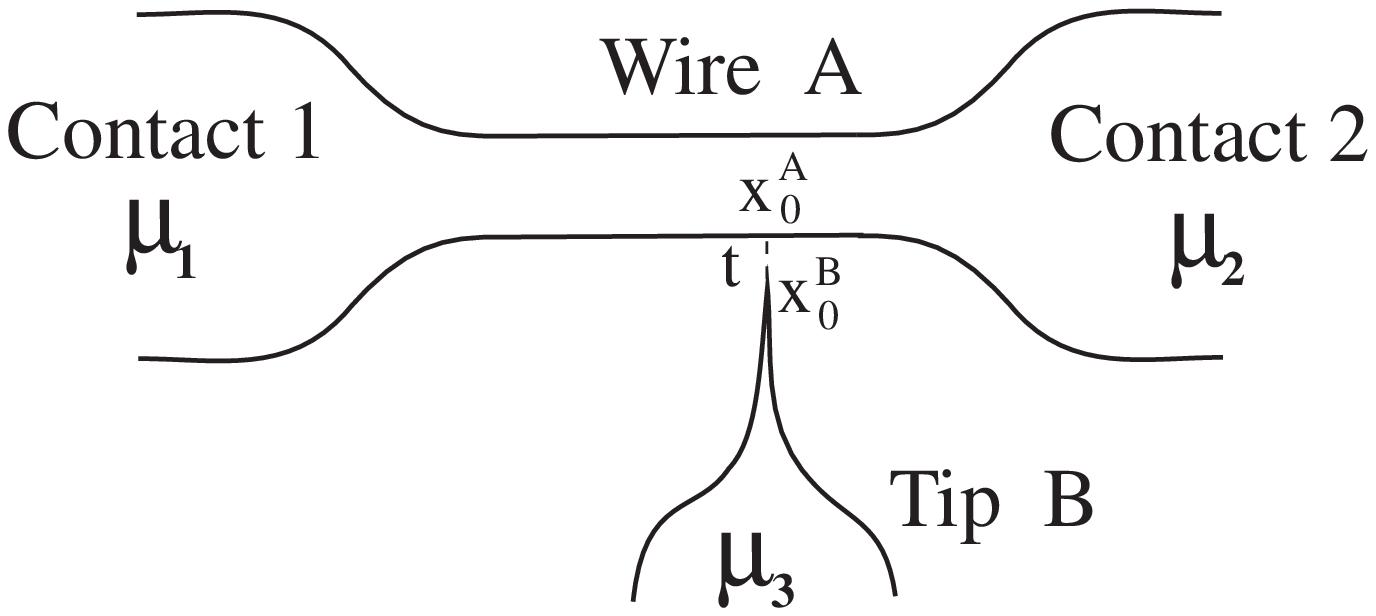}
\caption{Probing the voltage with an STM tip on a mesoscopic wire. Wire and tip
are coupled via a single bond connecting point $x_0^A$ in the wire to
point $x_0^B$ in the tip with a coupling element $t$.} \label{wirefig}
\end{figure}

Tunneling from a small metallic tip or from a suitable 
mesoscopic contact into a sample is a powerful means for the structural 
analysis of surfaces on an atomic length
scale \cite{rohrer85}.
In the typical arrangement 
a two-terminal measurement is performed using the tip (contact) as a 
source and the sample as a current sink.
Modeling the tip as a spherical symmetric object (s-wave) and
using Bardeen's approach\cite{bardeen61} the current flowing from the
tip into the sample
was found to be proportional to the local density of states (LDOS) of the
surface at the position of the tip\cite{tersoff85,hansma87}.
In this article we 
investigate arrangements in which the sample is so small that the phase
coherence length exceeds all sample dimensions. The sample is connected
to several contacts, which
makes it possible to investigate it in a transport state. 
In particular, multiterminal resistance measurements become possible. 
In one interesting configuration 
two of the contacts of the sample act as current 
source and sink and the STM (or contact) serves as a voltage probe
(see Fig. \ref{wirefig}).
Pioneering experiments using such a configuration have been undertaken
by Muralt et al.\cite{muralt87} and Kirtley et al.\cite{kirtley88}\ . 
Theoretically weak coupling contacts to small conductors were 
considered already by Engquist and Anderson\cite{engquist81} to find the
resistance of
a scatterer. The discussion of Engquist and Anderson led to Landauer's
result\cite{landauer} which expresses 
the resistance of a scatterer in terms of the ratio of its  
reflection probability $R$ and its the transmission probability $T$.
Landauer's derivation makes no appeal to measurement 
probes, but following Engquist and Anderson subsequent multichannel 
generalizations of this result by Azbel\cite{azbel81} and B\"uttiker
et al.\cite{buttiker85} were also given interpretations as weak coupling 
measurements by Imry\cite{imry86}. 
It was later pointed out that the discussion 
of Engquist and Anderson is not exact but neglects Friedel-like
oscillations, generated by the interference 
of incident waves reflected at the scatterer\cite{buttiker89}.
Interest in the multichannel generalization of Azbel faded when it 
was noticed that it could not account for the magnetic field asymmetry
observed in metallic diffusive wires\cite{stone85}. Furthermore,
in mesoscopic physics experiments probes typically are massive 
and cannot be treated as a weak coupling measurement. 
A general multiterminal approach which treats all probes on an equal footing 
was introduced by B\"uttiker\cite{buttiker86}. 
This approach, which expresses resistances as rational functions
of transmission probabilities, also permits a discussion of weak coupling
probes. But to our knowledge 
a detailed investigation of weak coupling probes based on 
this general approach has not been carried out. 
Below we will present such an analysis in which the weak coupling 
probes and the sample are described with {\it one} overall scattering matrix. 

While the initial experiments by Muralt et al.\ \cite{muralt87} and Kirtley
et al.\ \cite{kirtley88}
are difficult to interpret improvements in sample preparation 
techniques and in low temperature STM will hopefully lead 
to a resumption of such measurements. 
In two terminal configurations Friedel oscillations
in the equilibrium electron density  
near steps\cite{hasegawa93,avouris94}
and in atomic chorals\cite{crommie93} have been observed
and there is no principal reason why a similar resolution could not 
be achieved in the measurement of the transport state. 
A detailed discussion of potential oscillations near a barrier 
has been given by Chu and Sorbello\cite{chu90}. These authors 
also find a marked difference between the potential measured at 
the weak coupling probe and the electrostatic potential in the bulk of 
the sample. 

It is the purpose of 
this work to derive general weak coupling formulae for multiterminal 
conductors. Of particular interest is a formulation in which 
the dual role of the weak coupling contact, which can act either 
as a current injector (or sink) or as a voltage probe, appears 
in a manifestly reciprocal way. We start from the multiterminal 
formulation of Ref. \cite{buttiker86,buttiker88} which is valid for all types of 
contacts. 

\begin{figure}
\narrowtext
\epsfysize3cm
\epsffile{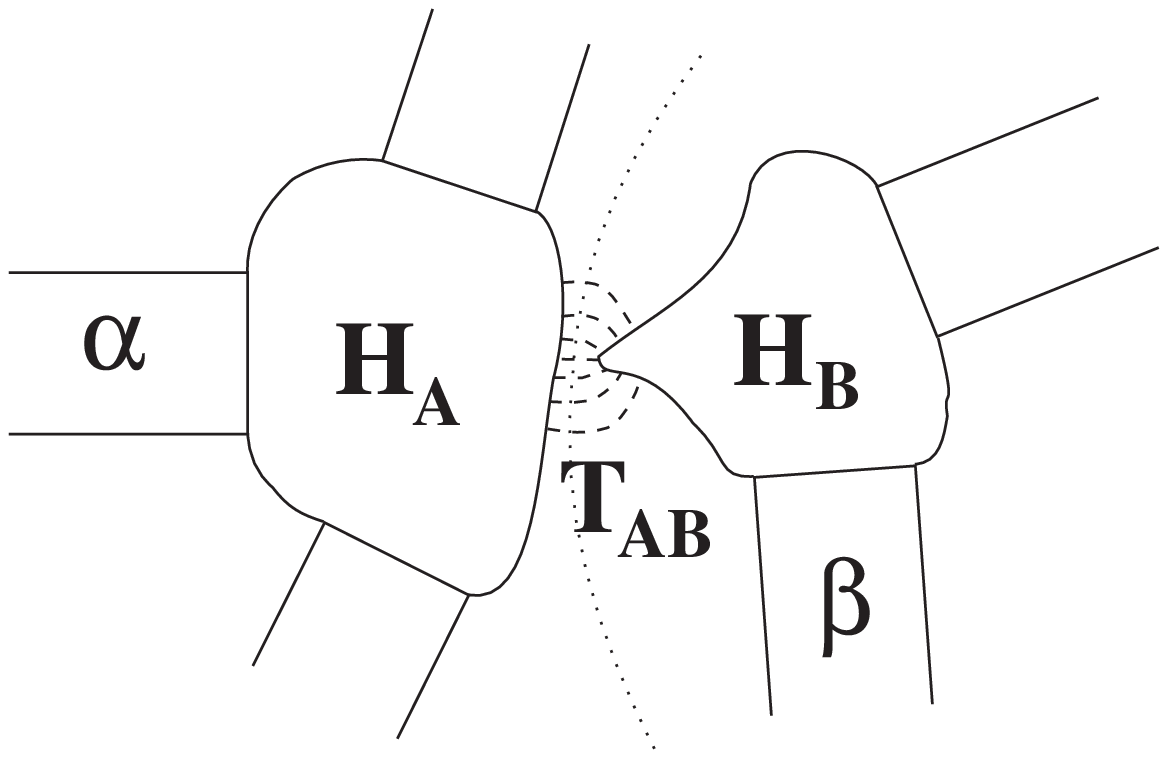}
\caption{Two mesoscopic multiterminal conductors (called system A and system B)
are weakly coupled by a coupling matrix ${\cal T}_{AB}$ which connects
a region of conductor A to a region of conductor B.}
\label{couplefig}
\end{figure}
The most general arrangement which we consider is 
depicted in Fig. \ref{couplefig}.
Two conductors $A$ and $B$ are weakly
coupled by a coupling matrix ${\cal T}_{AB}$.
Each conductor is connected via ideal leads to several electron reservoirs.
For the special case that the two conductors are only coupled by a single
tunneling path from $x^A_0$ to $x^B_0$ like it is the case in Fig.\
\ref{wirefig}
we find that the transmission probability
from contact $\alpha$ of conductor $A$ through the weak coupling contact 
to contact $\beta$ of conductor $B$ 
is given by a generalized Bardeen-like expression,
\begin{equation}
T_{\beta\alpha}=4\pi^2\nu_B(\beta, x^B_0)|t|^2\nu_A(x^A_0,\alpha)\, .
\label{multitran}
\end{equation}
Here $t$ is a coupling energy, $\nu_A(x^A_0,\alpha)$ is the injectivity
of contact $\alpha$ into point $x^A_0$ and $\nu_B(\beta,x^B_0)$ is the 
emissivity of point $x^B_0$ into contact $\beta$. The injectivities 
and emissivities\cite{buttiker93,gasp96} are only a portion of the LDOS
and will be denoted as local partial density of states (LPDOS). 
Below we give analytical expressions for these
densities of states in terms of functional derivatives of the 
scattering matrix of conductor $A$ and $B$ and in 
terms of wavefunctions \cite{buttiker93,buttchr96}.
Here we mention only that in the presence of a magnetic field $B$
the injectivity and emissivity in each conductor are related 
by a reciprocity relation: The injectivity from a contact $\alpha$
into a point $x$ in a magnetic field $B$ is equal to the 
emissivity of this point into contact $\alpha$ if the field is reversed, 
\begin{equation}
\nu( B; x, \alpha) = \nu(- B; \alpha, x)\, .
\label{magsym}
\end{equation}
Consequently the transmission probability given by Eq. (\ref{multitran})
manifestly obeys the Onsager-Casimir symmetry $T_{\alpha\beta} (B)
= T_{\beta\alpha} (-B)$. 
Eq.\ (\ref{multitran}) is one of the central results of this work. 
It can be used to find the resistances in an arbitrary 
multiterminal geometry with the help of formulae that 
express these resistances as rational functions 
of the transmission probabilities. 

Like transmission probabilities the LPDOS are evaluated in the 
equilibrium electrostatic potential. They are thus affected by interaction
(screening, etc.) only to the extend that interactions 
determine the equilibrium electrostatic potential. On the other hand 
the LPDOS are not sufficient to determine the actual electron
distribution in the sample. The change in the charge density in 
response to an increase of the electrochemical potential, say at 
contact $\alpha$ by $d\mu_{\alpha}$, is given by 
$dn(x) = \nu(x,\alpha)d\mu_{\alpha} +dn_{ind}(x)$
where $dn_{ind}(x)$ is the induced charge density generated 
by the non-equilibrium electrostatic potential \cite{buttiker93}. 

Below we introduce the Hamiltonian 
formulation\cite{verbaarschot17,iida90,lewenkopf91} of the scattering matrix
and express the LPDOS using this approach. The LPDOS are also expressed in
terms of scattering states.
We present the derivation of Eq. (\ref{multitran}) 
starting from the full scattering matrix of the weakly coupled system.
We then discuss a number of applications and the relation
of our results to the earlier work mentioned above. The magnetic field
dependence of our resistance formula is illustrated on a ballistic conductor
with a barrier.
\section{Hamiltonian formulation of the local partial densities of states}
In a first step we derive expressions for the local
partial density of states in terms of the Hamiltonian of the sample.
Let us consider a mesoscopic conductor which is connected via ideal leads
to $N$ electron reservoirs. We assume that, at the Fermi energy, 
we have in each lead $\alpha$, $N_\alpha$ open channels.
The Hamiltonian approach\cite{verbaarschot17,iida90,lewenkopf91}
starts with a formal
division of the Hilbert space into two parts, the open leads and the compact
sample region. The Hamiltonian can then be written as
\begin{equation}
{\cal H}=K+H+W+W^\dagger\, .
\label{ham}
\end{equation}
Here
\begin{equation}
K=\sum_{\alpha=1}^N \sum_{m=1}^{N_\alpha} |\alpha
m\rangle\langle \alpha m|E_F
\end{equation}
is the Hamiltonian of the isolated leads,
\begin{equation}
H=\sum_{x,x'} |x\rangle\langle x'| H_{xx'}
\end{equation}
is the Hamiltonian of the
isolated conductor and finally
\begin{equation}
W=\sum_x \sum_{\alpha=1}^N
\sum_{m=1}^{N_\alpha} |x\rangle\langle \alpha m| W_{x,\alpha m}
\end{equation}
describes
the coupling of the leads to the conductor. The set $\{|\alpha m\rangle\}$
represents a basis of scattering states in the isolated leads at the Fermi
energy $E_F$. The Hilbert space of the cavity
is spanned by $M$ localized states $|x\rangle$.  These two sets of states
form a complete basis of the Hilbert space of the total system.

%We make the assumption that the
%`` background'' scattering matrix $S_0$ is diagonal.  That means that without
%the coupling to the cavity an incoming weave in channel $m$ of lead $a$
%is only reflected into an outgoing weave in the same channel shifted by an
%arbitrary phase.  The matrix $S_0$, hopefully, does not play any role in our
%further discussion. 
%
The on-shell scattering-matrix for this
system at the energy $E_F$ is given by
\begin{equation}
S(E_F)=1-2\pi i W^\dagger GW\,
, \label{smatrix}
\end{equation}
with the Greens function
\begin{equation}
G=(E_F-H+i\pi WW^\dagger)^{-1}\, .  \label{denom}
\end{equation}
The matrix elements of $S$ can be written as
\begin{equation}
s_{\alpha m,\beta n}=\delta_{\alpha\beta}\delta_{mn}-2\pi i W_{\alpha
m}^\dagger G W_{\beta n}\, , \label{smelement}
\end{equation}
where we introduced partial coupling matrices
\begin{equation}
W_{\alpha m}=\sum_{x} |x\rangle\langle \alpha m| W_{x,\alpha m}\,
.
\end{equation}
These matrices describe the coupling of a single channel
$|\alpha m\rangle$ to the cavity.  With this definition we can decompose the
$M\times M$ matrix $WW^\dagger$ into a sum
\begin{equation}
WW^\dagger=\sum_{\alpha=1}^N
W_\alpha W_\alpha^\dagger =\sum_{\alpha=1}^N \sum_{m=1}^{N_\alpha} W_{\alpha
m}W_{\alpha m}^\dagger\, .
\end{equation}
For later use, we define
\begin{displaymath}
\Gamma_\alpha = W_\alpha W_\alpha^\dagger\qquad{\rm and}
\qquad \Gamma=\sum_\alpha \Gamma_\alpha\, .
\end{displaymath}
We want to express the LPDOS 
in terms of the Hamiltonian Eq. (\ref{ham}). 
This is possible with the help of expressions which relate the LPDOS
to functional derivatives of the scattering 
matrix\cite{buttiker93,gasp96,buttchr96}. 
The LPDOS of carriers which are injected through contact $\beta$,
reach point $x$ and are 
emitted into contact $\alpha$ is given by 
\begin{equation}
\nu(\alpha,x,\beta) = 
-\frac{1}{4\pi i}Tr\left[ s_{\alpha\beta}^\dagger \frac{\delta
s_{\alpha\beta}}{\delta U(x)} - h.c.\right] \, ,  \label{pardos}
\end{equation}
where $s_{\alpha\beta}$ is that submatrix of the full scattering matrix which
describes scattering from all the channels in contact $\beta$ into all the
channels in contact $\alpha$.
The injectivity of contact $\beta$ into point $x$ is the 
sum of all LPDOS over all contacts 
through which a carrier can possibly exit the sample,
\begin{equation}
\nu(x, \beta) = 
-\frac{1}{4\pi i}\sum_{\alpha} Tr\left[ s_{\alpha\beta}^\dagger \frac{\delta
s_{\alpha\beta}}{\delta U(x)} - h.c.\right] \, .  \label{injec}
\end{equation}
The emissivity into contact $\alpha$ of point $x$ is the 
sum of all LPDOS over all contacts through which a carrier 
can possibly enter the sample, 
\begin{equation}
\nu(\alpha,x ) = 
-\frac{1}{4\pi i}\sum_{\beta} Tr\left[ s_{\alpha\beta}^\dagger \frac{\delta
s_{\alpha\beta}}{\delta U(x)} - h.c.\right] \, .  \label{emiss}
\end{equation}
Finally, the local density of states is given 
by the sum over all injectivities,
or a sum over all emissivities, or the sum of all LPDOS,
\begin{equation}
\nu(x) = \sum_{\alpha}\nu(\alpha,x) =\sum_{\beta}\nu(x,\beta) =
\sum_{\alpha\beta}\nu(\alpha,x,\beta) . 
\end{equation}
To find
the functional derivation of $S$ with respect to $U(x)$, we notice that in the
discretized Hamiltonian the potential $U(x)$ appears only in the diagonal
terms $H_{xx}=E_x+U(x)$.  Therefore we can 
express the functional derivative with respect to the potential 
as an ordinary derivative with respect to the diagonal 
elements of the Hamiltonian\cite{brouwer97}, 
\begin{equation}
\frac{\delta
s_{\alpha\beta}}{\delta U(x)} = \frac{\delta s_{\alpha\beta}} 
{\delta H_{xx}}\,.
\end{equation}
Using Eq.\ (\ref{smelement}) for the $S$-matrix elements and 
Eq. (\ref{pardos}) we find
\begin{eqnarray}
\nu(\alpha,x,\beta) & = & Re\left(\delta_{\alpha\beta}\left[
G\Gamma_\alpha G \right]_{xx}\right. \nonumber \\
& + & \left. 2\pi i\left[G\Gamma_\beta G^\dagger \Gamma_\alpha
G\right]_{xx}\right)\, . 
\end{eqnarray}
Here, $[A]_{xx}$ denotes the diagonal
element of the $M\times M$ matrix $A$.
Taking into account that the $S$-matrix is unitary 
we get for the injectivity
\begin{eqnarray}
\nu(x,\beta) & = & \sum_\alpha
\nu(\alpha,x,\beta) =\left[G\Gamma_\beta G^\dagger 
\right]_{xx} \label{injglg}\\
& = & \int G(x,x_1)\Gamma_\beta(x_1,x_2)G^\dagger 
(x_2,x) dx_1 dx_2
\end{eqnarray}
and for the emissivity
\begin{equation}
\nu(\alpha,x)=\sum_\beta
\nu(\alpha,x,\beta) =\left[G^\dagger\Gamma_\alpha G
\right]_{xx}\, . \label{emiglg}
\end{equation}
Note that the injectivity depends in an explicit
manner only on the coupling elements 
$W_\beta$ to the leads through which carriers enter and 
the emissivity depends only on the coupling elements 
to the lead to which carriers finally exit. 

The $M\times M$ matrices 
\begin{eqnarray}
\underline{N}_{\alpha} & = & G\Gamma_\alpha G^\dagger\, ,
\label{injeop} \\
\overline{N}_{\alpha} & = & G^\dagger\Gamma_\alpha G \, 
\label{emisop}
\end{eqnarray}
whose diagonal elements are the injectivity 
(emissivity) are called injectivity (emissivity) operator.
However, the non-diagonal terms of these operators are not LDOS
but play an important role in the description of extended tunneling contacts. 
The total LDOS is given by 
\begin{equation}
\nu(x)=\sum_{\beta\alpha}
\nu(\alpha,x,\beta) =\left[G\Gamma
G^\dagger \right]_{xx}\, . \label{lodos}
\end{equation}
For later reference we 
also need the probability for transmission of a carrier from 
contact $\alpha$ into a different contact $\beta$. It is obtained by summing the
squared value of the $S$-matrix element $|s_{\beta n,\alpha m}|^2$ over all
channels $m$ of contact $\alpha$ and all channels $n$ of contact $\beta$,
\begin{eqnarray}
T_{\beta\alpha} & = & \sum_{m\in \alpha\atop n\in \beta} 
|s_{\beta n,\alpha m}|^2 \nonumber \\
 & = & 4\pi^2 Tr\left[G^\dagger\Gamma_\beta G \Gamma_\alpha
\right]\, .  \label{tba}
\end{eqnarray}
Eq. (\ref{tba}) is a standard result that is discussed in
textbooks\cite{datta}.
\section{Injectivity and emissivity and scattering states}
In this section we discuss briefly expressions for the injectivities
and emissivities in terms of the scattering states $\psi_{\alpha m} (x)$ for
a system described in the previous section.
A scattering state is obtained from 
a spatially very wide (energetically very narrow) wave packet 
which is incident in contact $\alpha$ in channel $m$ in the limit that 
the energy spread tends to zero. The scattering state
consists thus 
of an incident wave in channel $m$ of lead $\alpha$ and 
typically of waves that are reflected back into all channels 
of lead $\alpha$ and of waves that are transmitted 
into all channels of all the other leads \cite{buttiker88}.
We assume that the incident wave 
is normalized to unity. 
The scattering states of all channels and all leads together with 
possible bound (localized) states form a complete orthonormal set of states. 
To establish a relation between the injectivity and the scattering states
a connection between the functional derivatives of the scattering
matrix and the scattering states is needed. Such a connection was found in the 
analysis of tunneling times based on the local Larmor 
clock\cite{leavens}. For a more recent discussion we refer the reader to Refs. 
\cite{gasp96,buttchr96}. Here we state only the results.

The injectivity of a contact $\beta$ into a point $x$ is that part
of the LDOS which is contributed by the scattering states 
describing particles incident from
contact $\beta$,
\begin{equation}
\nu(B;x,\beta)=\sum_m\frac{1}{hv_{\beta m}(B)}|\psi_{\beta m}(B;x)|^2\, .
\label{winject}
\end{equation}
where $v_{\beta m} = \sqrt{2/m(E-E_{0,\beta m})}$
is the velocity of electrons with energy $E$ in channel $m$ of lead $\beta$
and $E_{0,\beta m}$ is the band offset.
$B$ is the magnetic field. Using the symmetry relation
(\ref{magsym}) the emissivity can be written
in terms of scattering states in the form
\begin{equation}
\nu(B;\alpha,x)=\sum_m\frac{1}{hv_{\alpha m}(-B)}|\psi_{\alpha m}
(-B;x)|^2\, .
\label{wemit}
\end{equation}
With the help of 
Eqs.\ (\ref{winject}) and (\ref{wemit}) we can describe the injectivity
and emissivity
on atomic length scales. The spatial variation contained in these densities 
of states is extremely complex not only inside a conductor but 
even in a perfect conductor leading up to a barrier. 
Consider a scatterer characterized by reflection amplitudes $r_{11,mn}$
for carriers incident in contact $1$ in channel $n$ and reflected into 
contact $1$ into channel $m$. In the lead connecting the scatterer to contact 1
the absolute squared value of the scattering state
$\psi_{1,n}$ has then "diagonal" contributions proportional to 
$R_{11,mn}\chi^{2}_{m}(y)$ which are spatially independent.
Here $R_{11,mn} = |r_{11,mn}|^{2}$ is the reflection probability
and $\chi^{2}_{m}(y)$ is the transverse wave function in channel $m$ in
lead $1$. 
In addition to these diagonal terms which are independent of $x$ 
there are interference terms proportional to $r^{\star}_{11,ln}r_{11,mn}
\chi_{l}(y)\chi_{m}(y)exp(i(k_{l}-k_{m})x)$ which oscillate on a very large
length scale 
for subbands with Fermi wave vectors which differ very little.
Suppose that we are not interested in the very detailed 
structure of the injectivity but only in the injectivity averaged 
over the cross section of the conductor.
Integration over $y$ eliminates the long range oscillations 
due to the orthogonality of the transverse wave functions. 
The only remaining oscillations along $x$ are then Friedel-like, proportional
to $\cos(2k_mx)$,
and the longest period is half a Fermi wavelength of the topmost 
occupied subband. Suppose that in addition to integrating over the 
transverse cross section we also average the density 
over a length large compared to this period. 
The resulting injectivity of contact $1$ to the left of the scatterer
is then 
\begin{equation}
\langle\nu(L,1)\rangle =
\sum\limits_{i=1}^{N_1}\frac{1}{hv_{i1}}+ \sum\limits_{i=1}^{N_1}
\frac{1}{hv_{i1}} R_{i}^{(11)}
\label{bilp1}
\end{equation}
and to the right of the scatterer is 
\begin{equation}
\langle\nu(R,1)\rangle = \sum\limits_{i=1}^{N_2}
\frac{1}{hv_{i2}} T_{i}^{(21)} .
\label{bilp2}
\end{equation}
Here the brackets $<>$ indicate the integration over $x$ and $y$.
The probabilities 
$R_{i}^{(11)} = \sum_{j} R_{11,ij} =  \sum_{j} |r_{11,ij}|^2$ and 
$T_{i}^{(21)} = \sum_{j} T_{21,ij} = \sum_{j} |t_{21,ij}|^2$ are the total
probabilities of all full incident channels which contribute to reflection
in channel $i$ and to transmission in channel $i$. 
Eqs.\ (\ref{bilp1}) and (\ref{bilp2}) will be useful to discuss the connection 
of the present work with the results of Azbel\cite{azbel81} and 
Refs. \cite{buttiker85},
and \cite{imry86}. We only note already at this point that the injectivities 
and emissivities averaged in this way cannot exhibit a full Hall effect since 
the Hall effect depends sensitively on the variation of the
injectivity and emissivity in the $y$ direction.  

We have now given expressions for the injectivities and emissivities
in terms of functional derivatives of the scattering matrices,
in terms of the diagonal elements of the injectivity and emissivity operators, 
and in terms of scattering states. Our next task is to express 
the transmission probabilities in terms of these densities of states. 
\section{Weak coupling of two conductors}
Consider now two conductors as shown in Fig. \ref{couplefig}.
Each of the conductors is described by a Hamiltonian 
${\cal H}_A$ and ${\cal H}_B$, defined according to
Eq.\ (\ref{ham}). The two conductors are coupled and this 
coupling is described by a matrix  
\begin{equation}
{\cal T}_{AB}=\sum_{x^A\in A \atop x^B \in B}|x^A\rangle\langle x^B|
t_{x^Ax^B}\, .
\label{coupmat}
\end{equation}
We are interested in the weak coupling limit and, therefore, 
take the matrix elements $t_{x^Ax^B}$ to be small.
The Hamiltonian for the total system reads
\begin{equation}
{\cal H}={\cal H}_A+{\cal H}_B+({\cal T}_{AB}+{\cal T}_{AB}^\dagger)\, .
\end{equation}
Let $G_A$ and $G_B$ be the Greens
functions defined in Eq. (\ref{denom}) of the uncoupled systems $A$ and $B$.
Then, to the
lowest order in $||{\cal T}_{AB}||$ we can write for the Greens function of the
coupled system,
\begin{equation}
G=G_A+G_B-G_A{\cal T}_{AB}G_B-G_B
{\cal T}_{AB}^\dagger G_A +{\cal O}(||{\cal T}_{AB}||^2)\, .
\end{equation}
We put this expression into Eq. (\ref{tba}) to
find the transmission probability of a contact $\alpha$ of system $A$ to a
contact $\beta$ of system $B$,
\begin{equation}
T_{\beta\alpha}=4\pi^2 Tr\left[{\cal T}_{AB}
G_B^\dagger \Gamma_\beta G_B{\cal T}_{AB}^\dagger
G_A\Gamma_\alpha G_A^\dagger\right]+{\cal
O}(||{\cal T}_{AB}||^4) \label{wtran}
\end{equation}
At first glance this formula looks quite complicated, but if one
compares it with Eqs.\ (\ref{injeop}) and (\ref{emisop}), 
one sees
that $T_{\alpha\beta}$ is just a combination of the injectivity- and
emissivity-operators and the coupling matrix ${\cal T}_{AB}$,
\begin{equation}
T_{\beta\alpha}=4\pi^2 Tr\left[{\cal T}_{AB}
\overline{N}_{\beta}(B) {\cal T}_{AB}^\dagger \underline{N}_{\alpha}(A) \right]
+{\cal O}(||{\cal T}_{AB}||^4)
\label{tgen}
\end{equation}
Let us now assume that the coupling of the two conductors
is point like. Then we can describe the coupling with a single weak bond. 
Thus, we set
${\cal T}_{AB}=|x^A_0\rangle\langle x^B_0|t$.
Putting this coupling matrix into Eq. (\ref{wtran}), the transmission
probability reduces to the simple expression given by Eq. (\ref{multitran}).
The probability of a carrier to go from contact 
$\alpha$ via the weak link 
to contact $\beta$ is given by the 
product of the injectivity of contact $\alpha$ into 
the connecting point $x^A_0$ multiplied by the 
emissivity of point $x^B_0$ into contact $\beta$. 

Suppose now that one of the conductors, for example conductor $B$, 
has only one contact. Then conductor $B$ provides a 
simple description of the tip of an STM. 
(Even though the current distribution between an STM tip and the surface is
spatially somewhat extended \cite{lang}, the coupling between tip and surface
is theoretically most often treated as being point-like.) 
If there is only a single contact the injectivity 
and the local density of states are 
identical, $\nu_B(x_0^B, \beta) \equiv \nu_B(x_0^B)$.
Thus the probability for transmission from a contact $\alpha$
of the sample into the tunneling tip is given by 
\begin{equation}
T_{\beta\alpha}=4\pi^2\nu_B(x^B_0)|t|^2\nu_A(x^A_0,\alpha)\, .
\label{tipsink}
\end{equation}
While on the tip-side only the local density of states enters,
on the sample side the relevant density of states is the 
injectivity. If the tip acts not as a carrier sink 
but as a carrier source the transmission probability from 
the tip into the sample is given by 
\begin{equation}
T_{\alpha\beta}=4\pi^2\nu_A(\alpha, x^A_0)|t|^2\nu_B(x^B_0)\, 
\label{tipsource}
\end{equation}
and contains on the sample side the emissivity as the relevant density 
of states. 

We see that the transmission is not proportional to the total LDOS
at the coupling point in the sample, but only to the injectivity 
(emissivity) of
that contact $\alpha$, for which we want to know the transmission probability
into the tip. This is due to the fact that the sample is connected
to more than one reservoir. If the sample is only connected to one electron
reservoir, all L(P)DOS of the sample are identical and
Eq.\ (\ref{multitran}) gives the Bardeen formula. The transmission
probability is $T =4\pi^2 \nu_A(x^A_0)|t|^2\nu_A(x^B_0)$. 
Thus, Eq.\ (\ref{multitran}) can be
seen as a generalization of the Bardeen formula for transmission between
two multiterminal conductors. Furthermore, Eq.\ (\ref{wtran}) is the
generalization of Eq.\ (\ref{multitran}) for the case when the tunneling
contact is
not pointlike, but allows for multiple tunneling paths. 
An equivalent expression for transmission between two one-terminal conductors
coupled via an extended tunneling contact has been given by Pendry et al.\
\cite{pendry91}.
\section{The voltage measurement}
Now we come back to the case, where the scanning tunneling
microscope is used to scan along a mesoscopic wire and 
to measure the voltage at
different points along the wire, c.\ f.\ Fig.\ \ref{wirefig}.
The electrochemical
potential which is applied to the tip reservoir 
is such that
there is no net current flowing through the tip.
At zero temperature, in terms of transmission probabilities, 
the measured electrochemical potential is given by\cite{buttiker88}
\begin{equation}
\mu_3=\frac{T_{31}\mu_1+T_{32}\mu_2}{T_{31}+T_{32}}\, \label{mu3form}
\end{equation}
in linear response to the
applied potentials $\mu_1$ and $\mu_2$. 
The transmission probabilities are evaluated 
at the Fermi energy, $T_{\alpha\beta}=T_{\alpha\beta}(E_F)$.
Putting our expressions for the transmission probabilities, Eq.
(\ref{multitran}), into this formula gives
\begin{equation}
\mu_3=\frac{\nu(x,1) \mu_1 + \nu(x,2) \mu_2}{\nu(x)} = 
\mu_2 + \frac{\nu(x,1)}{\nu(x)}(\mu_1-\mu_2)\, . \label{mu3mess}
\end{equation}
First, we remark that since the L(P)DOS depend on the
position $x$ in the wire where the tip is placed, also the measured
potential $\mu_3$ depends on this position.
Second, the measured potential does neither depend on the DOS
in the tip, nor on the coupling strength $t$ between tip and sample.
All terms in the nominator as well
as in the denominator are proportional to the coupling constant $t$ and
the density in the tip $\nu_B(x)$ so that these terms drop out.
(For non-zero temperature, if the DOS in the tip depends 
significantly on energy \cite{levyyeyati}, the measured voltage depends also on 
the density of states of the tip).

This formula allows us now to assign to every point
on the wire an electrochemical
potential. However, there is no simple relation between this measured
electrochemical potential and the electrostatic potential in the wire. The
electrostatic potential itself can not be measured, at least not using
the method described here. Measuring the electrochemical potential $\mu_3$
at a certain point $x$ does not mean that the electrons at that point $x$
locally are distributed according to a Fermi function with the electrochemical
potential $\mu_3$. We emphasize that there is no inelastic scattering inside
the sample. Applying a bias $\mu_2-\mu_1$ brings the system to a
non-equilibrium state so that the electrons inside the sample are not
distributed according to a Fermi distribution.
Next we illustrate the content of this formula with two examples. 
\subsection{Friedel-like oscillations across an impurity}
Eq.\ (\ref{mu3mess}) is valid for any distribution of impurities in the wire.
The only problem is to find expressions for the L(P)DOS. For a complicated
geometry with randomly distributed impurities there is no hope to find
an exact analytic expression for the L(P)DOS. But we can consider a simplified
example, which can give an idea how the measured potential should look
like in the neighborhood of an impurity.

\begin{figure}
\narrowtext
\epsfysize5cm
\epsffile{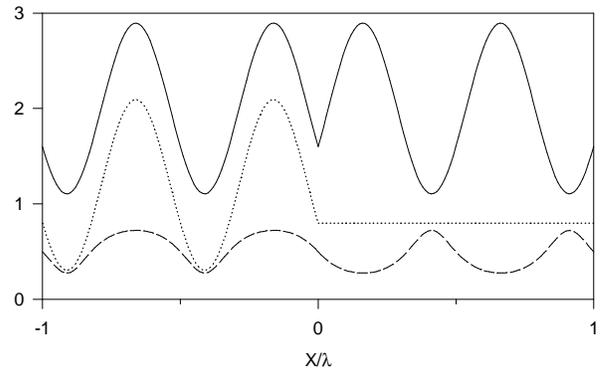}
\caption{The LDOS $\nu(x)$ (solid line), injectivity $\nu(x,1)$ 
of contact $1$ (dotted line) (both in units of $1/hv$), and the ratio of
these two densities
$\nu(x,1)/\nu(x)$ (dashed line) of a one dimensional wire with a $\delta$
barrier at $x=0$ which leads to a transmission probability of $T = 0.8$.}
\label{oscfig}
\end{figure}

We consider a one channel perfect conductor which contains only one
scatterer. Discussions closely related to the point of view taken here
are given by Levinson\cite{Levin} and in Ref. \cite{buttiker89}.
We assume that the equilibrium potential 
is constant all along the wire except for a delta-peak at $x=0$. On the
left side, the wire is connected to electron reservoir 1, and to the right side
to electron reservoir 2. 
For this model\cite{gasp96}, it is easy to find the analytic form of $\nu(x)$
and $\nu(x,1)$. 
In Fig.\ \ref{oscfig}
both quantities as well as their ratio are shown as functions of the
position $x$. The LDOS, which consists of contributions of scattering states
coming in from the left and from the right side, oscillates on both sides
of the
scatterer. The injectivity of contact one consisting only of contributions
of scattering states coming from the left side
oscillates only to the left of the
scatterer. Only there we have interference of incoming and reflected waves.
To the right of the scatterer, we have only outgoing waves so that the
injectivity of contact one is constant.
The $x$ dependence of the measured potential comes from the oscillations in
the ratio of the two densities, Eq.\ (\ref{mu3mess}). Since the injectivity is
a part of the LDOS, this ratio is always between zero and one and thus, the
measured potential $\mu_3$, Eq.\ \ref{mu3mess}, lies always between $\mu_1$
and $\mu_2$.
If the ratio is close to one, which is often the case to the left of the
scatterer, the measured potential $\mu_3$ is close to the applied
electrochemical potential to the left of the scatterer $\mu_1$. However,
we can also find positions to the left of the scatterer where the ratio
is small so that the measured potential is close to the electrochemical
potential to the right of the scatterer $\mu_2$. Likewise, we can also find
positions to the right of the scatterer where we measure a potential
which is close to the applied potential at the left side.
This leads to interesting effects when the voltage probe is used to measure
the resistance of a barrier.
\section{The Resistance Measurement}
\begin{figure}
\narrowtext
\epsfysize3cm
\epsffile{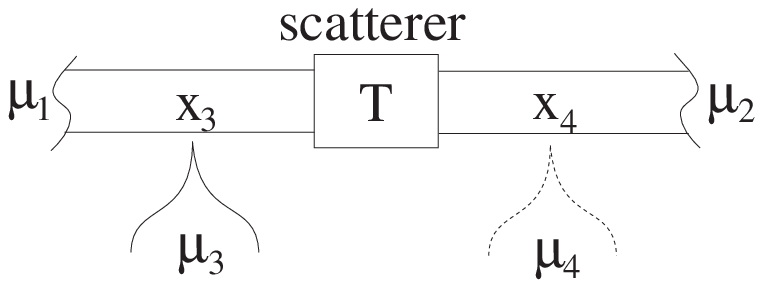}
\caption{Experimental setup for the measurement of the four-terminal resistance
of a scatterer using an STM tip as a voltage probe.}
\label{fourtermres}
\end{figure}
The four-terminal resistance\cite{buttiker88} (Fig. \ref{fourtermres}) of a scatterer 
%characterised by a scattering matrix
%
%\begin{equation}
%{\bf S}=\left( \begin{array}{cc} {\bf s}_{11} & {\bf s}_{12} \\
%{\bf s}_{21} & {\bf s}_{22} \end{array}
%\right)
%\end{equation}
%
is defined as the ratio
of the voltage drop across the scatterer, $\mu_3-\mu_4$, divided by the
current flowing through the scatterer due to the applied voltage
$\mu_1-\mu_2$,
\begin{eqnarray}
{\cal R}_{12,34} & = & \frac{\mu_3-\mu_4}{eI_{12}} \\ 
 & = & \frac{h}{e^2} \frac{1}{T}
\frac{(T_{31}T_{42} - T_{32}T_{41})}
{(T_{31} + T_{32}) (T_{41} + T_{42})}  
\label{fourt}
\end{eqnarray}
Here 
$T=Tr({\bf s}_{12}{\bf s}_{12}^\dagger)$ is the overall
transmission probability of the conductor
in the absence of the measuring contacts. Using Eq.\ (\ref{multitran})
the four terminal
resistance is 
\begin{eqnarray}
{\cal R}_{12,34} & = & \frac{h}{e^2} \frac{1}{T} 
\left(\frac{(\nu({\bf x}_3,1) \nu({\bf x}_4,2) - \nu({\bf x}_3,2) \nu({\bf x}_4,1)}
{\nu({\bf x}_3) \nu({\bf x}_4)}\right) \\
& = & \frac{h}{e^2} \frac{1}{T}
\left(\frac{\nu({\bf x}_3,1)}{ \nu({\bf x}_3)}
- \frac{\nu({\bf x}_4,1)}{ \nu({\bf x}_4)}
\right) \, . 
\label{fournu}
\end{eqnarray}
Notice that the last expression is just the difference of two 
(three terminal) voltage measurements given by Eq. (\ref{mu3mess}). 
The notion ${\bf x}_{i}=(x_{i},y_{i})$,$i = 3, 4$ denotes the coupling points
of the voltage probes 3 and 4 in the $x-y$ plane of the wire.
Since in the densities $\nu({\bf x},\alpha)$ and $\nu({\bf x})$
interference between incoming and reflected waves is taken into account,
Eq.\ (\ref{phsenres}) is a phase-sensitive resistance.
Due to these interference effects the densities show a complicated spatial
behavior, e.\ g.\ the Friedel-like oscillations in the one-channel case.

Instead of using the weak coupling contacts 
as voltage contacts we might also use contact 3 to inject current and 
contact four as the current sink. The measured resistance is then 
${\cal R}_{34,12}$ and is related to  ${\cal R}_{12,34}$
by a reciprocity relation\cite{buttiker86,buttiker88}  
${\cal R}_{34,12}(B) = {\cal R}_{12,34} (-B)$.
Thus this resistance is determined by the difference of the {\it emissivities}
into contact $1$ of the points ${\bf x}_3$ and ${\bf x}_4$,
\begin{eqnarray}
{\cal R}_{34,12} 
& = & \frac{h}{e^2}\frac{1}{T}
\left(\frac{\nu(1,{\bf x}_3)}{ \nu({\bf x}_3)} - \frac{\nu(1,{\bf x}_4)}
{\nu({\bf x}_4)}
\right) \, . 
\label{phsenres}
\end{eqnarray}
Note that both the overall transmission probability $T$ and the 
local densities $\nu({\bf x})$ are {\it even} functions of the magnetic field. 
We also remark that one might believe that current transport from one 
weak coupling probe to another invokes more information then is contained
in the injectivities or emissivities. This not the case, since the
current balance
is such that to lowest order in the coupling strength $|t|^2$
 of the weak coupling
contacts the injected current first reaches the massive contacts $1$ and $2$
and the current at probe $4$ is determined by carriers injected by the massive 
contacts $1$ and $2$. Direct transmission of carriers from one weakly coupled
contact to the other one is a second order effect, proportional to $|t|^4$,
and thus, generally only a small perturbation. Chan and Heller showed
recently\cite{heller97} that, on surfaces with point defects (adatoms), even
this second order effect can be deduced
from single tip measurements.

In order to get from Eq.\ (\ref{fournu}) to a
position independent value for the four-terminal resistance we average
the phase-sensitive result by moving the voltage probes on both sides
of the scatterer over some distance while measuring the voltage.
One possibility is to keep the transverse
coordinates $y_\alpha$ fixed and average only along the $x$ axis.

Let us now consider the one-channel case. There we know that it is sufficient
to average over half a Fermi wavelength, since the measured voltages show a
periodic behaviour on this length scale. 
We get
the phase-averaged result,
\begin{eqnarray}
\langle{\cal R}_{12,34}\rangle & = & \frac{h}{e^2}\frac{1}{T}
\left(\left<\frac{\nu(x_3,1)}{\nu(x_3)}\right>-
\left<\frac{\nu(x_4,1)}{\nu(x_4)}\right>\right) \label{phaver} \\
& = & \frac{h}{e^2}\frac{1-\sqrt{T}}{T}\, .
\end{eqnarray}
The same result was already found by B\"uttiker\cite{buttiker89} who
described the voltage probes as wave splitters.

For comparison we calculate also a phase-insensitive resistance which means
that we neglect the phase coherence of incoming and reflected wave
altogether. This is equivalent to averaging injectivity and local
density separately and leads to the Landauer\cite{landauer} formula for 
the resistance of a scatterer,
\begin{eqnarray}
{\cal R}_{12,34}^{insens} & = & \frac{h}{e^2}\frac{1}{T}\left(
\frac{\langle\nu(x_3,1)\rangle}{\langle\nu(x_3)\rangle}
-\frac{\langle\nu(x_4,1)\rangle}{\langle\nu(x_4)\rangle}\right)
\label{phinsenres} \\
& = & \frac{h}{e^2}\frac{1-T}{T}\, .
\end{eqnarray}
We emphasize that what can be measured directly by using a sufficiently sharp
tip is the phase-sensitive result. By moving the tip
and averaging over the measured potentials we get the phase-averaged
result. Note that the phase-averaged result would also be obtained 
if the measurement is made further then a phase-breaking length away 
from the scatterer.  
\subsection{The few-channel case}
For a scatterer connected to leads with $N>1$ open channels
Eq.\ (\ref{phsenres})
is still valid.
The injectivity as well as the
local density are now not anymore periodic functions. They consist of a
superposition
of oscillations with different wavelengths. In fact, the functions
contain oscillations with wave vectors $k=k_i\pm k_j$ given by all possible
combinations of the Fermi wave vectors $k_i$ of the $N$ channels.
If the number of channels is very large, one expects that the densities become
nearly constant as a function of the position $x$. 
Instead of the exact densities we can then use 
the densities averaged over a portion of the conductor. 
The injectivity of contact 1 to the left and to the right
of the scatterer averaged over $x$ and $y$
are given by Eqs.\ (\ref{bilp1}) and (\ref{bilp2}).
Using these densities in Eq.\ (\ref{phinsenres}) 
gives the result of Azbel\cite{azbel81} and 
B\"uttiker et al.\ \cite{buttiker85} found with the help of 
a charge neutrality argument.

Since the densities averaged over the $x$ and $y$ coordinates do not 
anymore exhibit a dependence on the transverse $y$ coordinate, 
the resulting resistance
formula can not explain the magnetic field dependence (Hall effect) of the
measured resistance. In contrast, Eq.\
(\ref{phsenres}), includes
the exact, spatial densities and 
shows a dependence on the magnetic field not only through
transmission probabilities but also due to the magnetic field
dependence of the injectivity and emissivity.  
We will now illustrate this point by investigating the 
Hall resistance of a one channel wire with an obstacle. 
\subsection{Magnetic field dependence of the Resistance}
Let us consider a scatterer which is connected via ideal
leads to two electron reservoirs. Let us assume that in the ideal lead,
far away from the scatterer, 
we have a
uniform potential in the longitudinal
$x$ direction and a parabolic confining potential in the transverse $y$
direction, $U(y)=1/2m\omega_0^2y^2$.
Furthermore, the lead is threaded by a magnetic field perpendicular
to the $x-y$ plane. In the lead
the eigenfunctions of such a system can be written
as a product of a plane wave $\exp(ikx)$ in $x$ direction and a transverse
wave function $\chi_k(y)$. The transverse wavefunction depends now on the
$k$ vector of the plane wave in the $x$ direction. That means, that not
only different channels, but also
incoming and reflected waves in the same channel have different transverse
wave functions. This mechanism leads to a spatial separation of incoming
and reflected waves and in a strong magnetic field to the formation of edge
channels\cite{streda,buhall}. The Hall effect in perfect ballistic wires
has been discussed\cite{peet} in connection with the suppression of the 
Hall effect in ballistic crosses\cite{roukes}. This suppression is, 
however, an effect which depends on the geometry of the cross\cite{ford,been}.
Here, the main effect which we investigate arises due to the scatterer
in the wire which is taken to have a magnetic field independent 
transmission probability $T = 0.5$.  

For the case of only one open channel in the lead, the scattering state
coming in from contact 1 can be written in the lead connecting the scatterer
to contact 1 as
\begin{equation}
\psi_1({\bf x})=e^{ik_+x}\chi_+(y) + r_{11}e^{-ik_-x}\chi_-(y)\, .
\label{magscatstate}
\end{equation}
Here, $r_{11}$ is the reflection amplitude for reflection from contact 1 into
itself, $k_+$ and $k_-$ are the wave vectors for incoming and reflected waves
and $\chi_+$ and $\chi_-$ are the corresponding transverse
wave functions\cite{datta}.
The voltage $\mu_3$ measured on a point to the left of the
scatterer is given by Eq.\ (\ref{mu3mess}). 
Using Eq.\ (\ref{magscatstate})
we find for the densities,
\begin{eqnarray}
\nu({\bf x},1) & = & \frac{1}{hv}(|\chi_+(y)|^2+|r_{11}|^2|\chi_-(y)|^2
\nonumber \\
& & +2|r_{11}\chi_+(y)\chi_-(y)|\cos(2kx+\delta))\, , \\
\nu({\bf x},2) & = & \frac{1}{hv}(1-|r_{11}|^2)|\chi_-(y)|^2\, , \\
\nu({\bf x}) & = & \nu({\bf x},1)+\nu({\bf x},2)\, .
\end{eqnarray}
Here, $k=(k_++k_-)/2$ and $r_{11}=|r_{11}|\exp(i\delta)$. The magnetic field
dependence of this formula is hidden in the transverse wavefunctions.
\begin{figure}
\epsfysize6cm
\epsffile{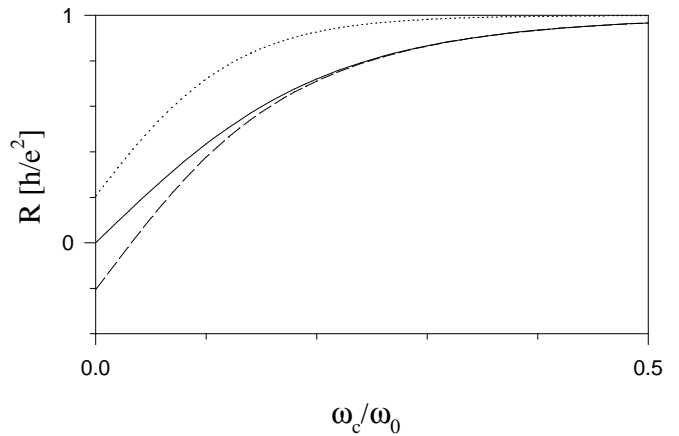}
\caption{Magnetic field dependence of the resistance
of a ballistic wire with a scatterer with transmission 
probability $T = 0.5$.  The voltage probes are
placed on opposite edges of the lead having the same $x$ coordinate (solid
line), being separated in $x$ direction by a quarter of a Fermi wavelength
(dotted line) and (dashed line). \label{magdep}}
\end{figure} 
In Fig.\ \ref{magdep} we show the measured four-terminal
resistance as a function of the ratio $\omega_c/\omega_0$, where $\omega_c
=eB/m$ is the cyclotron frequency and $B$ the magnetic field.
The three curves correspond to three different
configurations of the voltage probes, contact 3 and 4.
The two voltage probes are placed on the two edges of the
lead connecting contact one to the scatterer.
First in a way that the $x$ position of the two probes is the same
(solid line). Second,
the $x$ coordinate of probe 3 is such that the LDOS has a minimum,
$\cos(2kx+\delta)=-1$, at that point in the lead, whereas probe 4 is place
on a point
where the LDOS has a maximum, $\cos(2kx+\delta)=1,$ (dashed line). 
And finally, probe 4 is
placed over a minimum of the LDOS and probe 3 is placed over a maximum of the
LDOS (dotted line). 
Note, that, if there is no magnetic field present, the resistance
is zero, when the positions of the two probes differ only in their $y$
coordinate. If the magnetic field is turned on, a Hall-resistance develops
which at strong magnetic fields reaches the 
quantized value ${\cal R}=h/e^2$.
In the case, where the positions of the probes differ in the $x$ coordinate
in the way described above, a non-vanishing resistance is already measured
at zero magnetic field. However, when the field is turned on, both
curves approach the value $h/e^2$.
For small magnetic fields, we can expand the resistance formula, Eq.\ 
(\ref{fournu}), and get 
to first order in $\omega_c/\omega_0$
\begin{eqnarray}
{\cal R}_{12,34} & = & \frac{h}{e^2}\frac{1}{T}\left(
\frac{\nu_0(x_4,2)}{\nu_0(x_4)}\right)-\frac{\nu_0(x_3,2)}{\nu_0(x_3)}\\
& & -\frac{\omega_c}{\omega_0}\left(
\frac{y_4}{y_0}\frac{\nu_0(x_4,2)}{\nu_0(x_4)} 
-\frac{y_3}{y_0}\frac{\nu_0(x_3,2)}{\nu_0(x_3)}
\right)\, ,
\end{eqnarray}
where $\omega_c = eB/m$ is the cyclotron frequency and $y_0=\sqrt{\hbar/
(m\omega_0)}$. In this formula, the $y$ dependent Hall resistance and the
$x$ dependent, longitudinal resistance enter additively.
\section{Discussion}
In this work we described systems consisting of two weakly-coupled multi-probe
conductors starting
from the global scattering matrix of the whole system
which covers all parts including the weakly coupled contact.
We derived a general transmission formula, Eq. (\ref{tgen}), for transmission
through the weak-coupling contact.
We have investigated this formula in the case where there is only 
one tunneling path. Using these expressions we can rewrite formulae, which
express the resistance as functions of transmission probabilities, as
functions of the local partial density of states.
Applying this result the resistance of a one-channel conductor with a barrier
shows interesting and surprising features.
Our resistance formula can also account for the Hall-resistance. This point
is illustrated using two weakly-coupled voltage-probes to measure the
magnetic-field dependence of the resistance of a ballistic one-channel
conductor with a barrier.

Based on the general expression, we can also treat contacts 
which permit multiple tunneling paths. 
After all, even an STM exhibits a current distribution with a 
certain spatial width\cite{lang}. It is then interesting to ask 
what the densities are which are measured by spatially extended
tunneling contacts. To our knowledge, a detailed study of extended tunneling
contacts has not yet been done.

We have treated the zero-temperature limit. 
At a non-vanishing temperature the corresponding results 
are obtained by multiplying the transmission probability with 
the Fermi function of the injecting reservoir. Even in the limit of a single 
tunneling path the corresponding resistances will then in general 
not be independent of the density of states in the tip 
if this density exhibits a substantial variation at the Fermi 
energy\cite{levyyeyati}. 

We have already emphasized 
that the density of states discussed here are quantities 
which are conjugate to the electrochemical
potential of a contact\cite{buttiker93}. At zero temperature they
are evaluated at the Fermi energy in the equilibrium potential.  
The densities used here are thus essentially chemical quantities 
and interactions enter only through the equilibrium potential. 
The interaction induced portion of the density does not
enter into the transmission behaviour of a conductor. 
This should be contrasted with the notion of tunneling-density of states 
which are evaluated using the unrestricted Green's functions containing
the full interaction. An investigation of this important 
point is beyond the scope of this work. We refer an interested 
reader to a discussion of the same issue concerning not density of 
states but directly the conductance of interacting systems\cite{alekseev}. 

We have discussed examples of voltage 
and resistance measurements on conductors with a single 
open channel. The approach discussed here, however, is also suitable for
conductors having several or even a large number of open channels. 
For a metallic diffusive conductor of length $L$, 
extending from $x =0$ to $x =L$,  
with a local density of states $\nu$ the injectivity 
can be separated into an ensemble averaged part given 
by $\nu (1-x/L)$ and a fluctuating part.
The average behavior gives us the 
linear voltage drop and the ohmic length dependence 
which we expect for such conductors. 
More interesting is the investigation of the Hall conductance:
although the local, ensemble averaged density of states of a metallic
conductor  
is independent of the magnetic field, the injectivities exhibit
a linear dependence on the magnetic field and 
and this gives rise to the Hall resistance, similar to the one-dimensional 
example discussed in Section VI. 

To conclude we emphasize that the investigation of tunneling contacts 
on mesoscopic conductors is an interesting subject which has 
so far found only limited attention.

This work was supported by the Swiss National Science Foundation.
\end{multicols}
\end{document}